\documentclass{elsart}

\usepackage{epsfig}
\usepackage{amssymb}
\usepackage{rotating}
\begin{document}

\begin{frontmatter}



\title{Correlating Strangeness enhancement and $J/\psi$ suppression in Heavy Ion collisions at $\sqrt s_{\rm NN} = 17.2$ GeV}


\author{F. Becattini}
\address{Universit\`a di Firenze and INFN Sezione di Firenze, Italy}
\ead{becattini@fi.infn.it}
\author{L. Maiani}
\address{Universit\`{a} di Roma `La Sapienza' and INFN Sezione di Roma 1, 
Italy}
\ead{luciano.maiani@roma1.infn.it}
\author{F. Piccinini}
\address{INFN Sezione di Pavia and Universit\`a di Pavia, Italy} 
\ead{fulvio.piccinini@pv.infn.it}
\author{A.D. Polosa}
\address{Universit\`{a} di Bari and INFN Sezione di Bari, Italy}
\ead{antonio.polosa@cern.ch}
\author{V. Riquer}
\address{INFN Sezione di Roma 1, Italy}
\ead{veronica.riquer@cern.ch}

\begin{abstract}
It is shown that the strangeness 
enhancement and the $J/\psi$ anomalous suppression patterns observed in heavy ion collisions at top SPS energy, $\sqrt s_{\rm NN} = 17.2$ GeV, exhibit an interesting correlation if
studied as a function of the transverse size of the interaction
region. The onset of both phenomena seems to occur when the size exceeds
$\approx 4$ fm. Strangeness enhancement  is defined in terms  of the strangeness undersaturation factor $\gamma_S$ and $J/\psi$ anomalous suppression in terms of the deviation from the absorption expected in a purely hadronic scenario.
\end{abstract}

\end{frontmatter}

\section{Introduction}
\label{intro}

The heavy ion programme at the CERN SPS was aimed at detecting signals for 
deconfined quark and gluon matter. Among the proposed signals, we shall 
consider here the enhancement of strange particle production, 
proposed in Ref.~\cite{Rafelski} and observed by NA49 and NA57, and the anomalous absorption of J/$\psi$ with respect to Drell-Yan 
pairs observed by NA50 and NA60 Collaborations, following the theoretical suggestions in Ref.~\cite{Satz}.

Until now, the two signals have been separately analysed, 
each of them giving  circumstancial evidence for the production of a 
deconfined phase of hadronic matter. More conclusive evidence would be 
provided by a correlation between them. Indeed, one expects that the melting 
of charmonia, as function of a suitable critical parameter, occurs 
{\it simultaneously} to the reduction of the strange quark mass, which makes 
it easier for strange particles to reach thermal equilibrium and to appear 
abundantly among the hadrons formed at chemical freeze-out.


\begin{figure}[ht]
\begin{center}
\epsfig{
height=7truecm, width=7.5truecm,
        figure=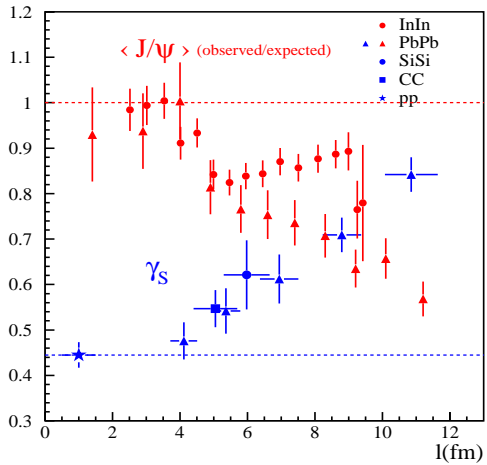}
\caption{\label{fig.1} \footnotesize Strangeness undersaturation
factor, $\gamma_S$,  and $J/\psi$ anomalous suppression ratio, $R_{J/\psi}$, as functions of the transverse size of the interaction region in heavy ion collisions at $\sqrt{s}_{\rm NN}=17.2$~GeV (see text for definitions). $J/\psi$ points are from Pb-Pb (NA50) and In-In (NA60) collisions (triangles and circles, respectively, with vertical error bars only); values of $\gamma_S$ refer to Pb-Pb, Si-Si, C-C and p-p collisions (triangles, circle, box, star, with vertical and horizontal error bars), data from NA49 and NA57.}
\end{center}
\end{figure}
In this Letter we would like to point out that indeed strangeness enhancement and $J/\psi$ suppression are strongly correlated when examined as 
function of the variable $l$ introduced in Ref.~\cite{MPPR}, related to 
the centrality of the collision.
With reference to the beam axis, $l$ is the transverse size of the overlapping 
region of the two colliding nuclei:
 \begin{equation}
l=2R-b,
\label{elle}
\end{equation}
$R$ is the radius of the nuclei at rest (we consider identical nuclei) and $b$ 
the impact parameter.

To characterize strangeness production, we consider the strangeness undersaturation
factor, or extra strangeness suppression factor, $\gamma_S$, used by many authors 
in the statistical analysis of hadron abundances. Values of $\gamma_S$ below unity 
mean that strange hadrons have not reached full chemical equilibrium in the hadron
resonance gas. Thus, $\gamma_S$ measures how difficult is for strange hadrons to
attain equilibrium starting from the initial collision of non-strange nuclear matter.

Our main result is shown in Fig.~1, where $\gamma_S$ is reported as function of $l$, together with the double ratio:
\begin{equation}
R_{J/\psi}=\left(\frac{Rate(J/\psi)/(D-Y)_{\rm Observed}}
{Rate(J/\psi)/(D-Y)_{\rm Expected}}\right).
\label{errePsi}
\end{equation}
Indeed, $\gamma_S$  departs from its low value in proton-proton or light nuclei collisions to approach unity in the same centrality range where $R_{J/\psi}$ drops below unity.
The possibility of such a correlation has been suggested in~\cite{catania}. An analysis of strange particle production versus centrality was given in~\cite{prl} but no correlation with $J/\psi$ production was considered.

Our considerations are phenomenological in essence, however we stress that $l$ is a good candidate to regulate the onset of the critical behaviour. The transverse lenght determines the linear size of the initial fireball and therefore its initial volume. It is most reasonable that collective phenomena like the ones we are addressing can take place only when the hadron matter involved reaches a certain minimum volume. In addition, the number of nucleons per unit surface which take part in the collision increases with $l$. Correspondingly, the energy density deposited in the fireball~\cite{Bj} and its initial temperature both increase so that, by varying $l$, we make a temperature scan which may lead into the critical or cross-over region.


\section{$J/\psi$ absorption vs. centrality}
\label{}

In Fig.~2,  the observed $J/\psi$ rate, normalized 
to Drell-Yan pairs, is reported as a function of $l$ for Pb-Pb \cite{NA50} and In-In \cite{NA60} collisions. 
The $J/\psi$ attenuation function is defined by~\cite{MPPR}:
\begin{equation}\label{abstot}
A(l)= \exp (-\rho_{{\rm nucl}}\sigma_{{\rm nucl}} L(l)) \
\exp\left (-\Sigma_i \langle \rho_i\sigma_i\rangle_T\cdot \frac{3}{8} l \right),
\end{equation}
%
%
\begin{figure}[ht]
\begin{center}
\epsfig{
height=7truecm, width=7.5truecm,
        figure=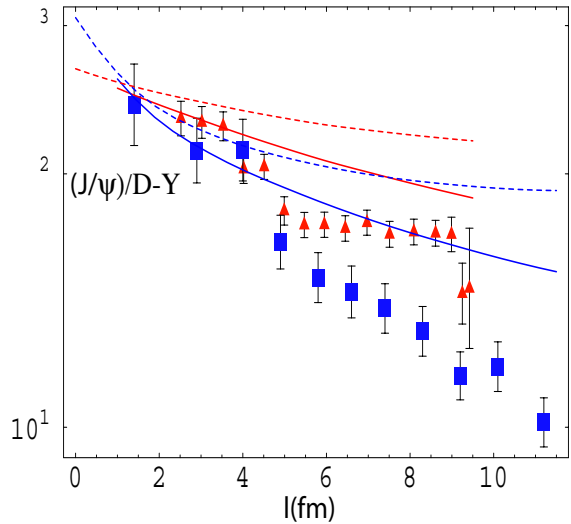}
\caption{\label{fig.1} \footnotesize $J/\psi$ production, normalized to Drell-Yan pairs, vs $l$ in Pb-Pb (boxes, NA50) and In-In (triangles, NA60) collisions. Dashed lines represent the expected values after nuclear absorption correction~\cite{NA60}; solid lines include the hadronic absorption, computed assuming a limiting Hagedorn temperature: $T_{Hag}=180$~MeV~\cite{MPPR} (upper = In, lower = Pb, for both dashed and solid lines).}
\end{center}
\end{figure}
$\rho_{\rm nucl}$ and $\rho_i$ are the number 
densities of nuclear matter and of the hadronic species $i$ in the fireball, respectively; correspondingly, the cross sections 
$\sigma_{{\rm nucl}}$ and $\sigma_{{\rm i}}$ are the nucleon-$J/\psi$ and 
hadron-$J/\psi$ absorption cross sections. A thermal average 
is taken for the latter at temperature $T$~\cite{MPPR}. 
The length $L$ is the average linear longitudinal size of the overlapping 
region of the two colliding nuclei. The values of $L$ as a function of the impact 
parameter $b$, for Pb and In, have been provided to us by the NA50 and NA60 collaborations~\cite{ram&scomp} and we have derived the corresponding values of $l$ with Eq.(\ref{elle}) and $R=6.5, 5.3$~fm, for Pb and In respectively. Finally, $3/8\cdot l$ is the average lenght traversed by a $J/\psi$ produced at random inside a sphere of diameter $l$.

The dashed curves in Fig.~2 give the expected rates after nuclear absorption 
corrections as computed by NA60~\cite{NA60}. Solid curves include the hadronic absorption corrections, see Eq.(\ref{abstot}),  (upper = In, lower = Pb for both dashed and solid lines). We outline here a few elements of the calculation of hadronic absorption given in Ref.~\cite{MPPR}, and refer the reader to the original paper for details.

Pseudoscalar and vector mesons cross sections for: 
\begin{equation}\label{reaction}
(\pi,\rho,{\rm K}^{(*)},\phi,\ldots)+J/\psi \to {\rm open\;charm}
\end{equation} 
have been computed in the Constituent Quark Model \cite{CQM}. The absorption length has been evaluated in a hadron gas at a temperature 
$T=175$~MeV which was adjusted to reproduce the data points for $l<4$~fm.The meson chemical potentials were neglected as a reasonable first-order approximation. 
%
%
Extrapolation to larger centralities requires the hadron gas Equation of State (EoS), 
to convert the energy density into temperature. We have used the EoS of a Hagedorn gas 
with limiting temperature $T_{H}=180$~MeV\cite{MPPR}. 
The ratio Observed (experimental points) to Expected (solid curves) is plotted vs. 
$l$ in Fig.~1. 

\section{Strangeness production}
\label{}

Primary strange particle densities in phase-space are parameterized according to:
\begin{equation}\label{gammas}
\rho_S(E,T)=(2J+1)(\gamma_S)^{n_S}\;exp\left ( \frac{{\vec \mu}\cdot {\vec q}-E}{T}\right),
\end{equation}
where $n_S$ is the number of valence strange quarks and antiquarks in the particle, ${\vec q}$ are a set of conserved charges, ${\vec \mu}$ the corresponding chemical potentials and $\gamma_S$ is the strange particle suppression factor. In order to obtain the parameters of the hadronic source at chemical freeze-out, i.e., temperature $T$, baryon-chemical potential $\mu_B$ and $\gamma_S$, we have fitted full-phase-space multiplicities
to those measured in peripheral Pb--Pb collisions and central C--C, Si--Si and 
Pb--Pb collisions at maximum SPS energy, $\sqrt s_{NN} = 17.2$ GeV. 
The fitting procedure is described in detail elsewhere \cite{bkgms}; the only difference 
with respect to Ref.~\cite{bkgms} is that we have used an 
updated hadronic data set \cite{pdg}. A more detailed report on these fits is the subject of a separate 
publication \cite{becman}. 

For central C--C and Si--Si we have used the average multiplicities of $\pi^{\pm}$, 
K$^\pm$, $\phi$ and $\Lambda$'s~\cite{prl,ccsisi} measured by the experiment NA49. 
For central Pb--Pb the data sample by NA49 includes
$\pi^{\pm}$ \cite{piK}, K$^\pm$ \cite{piK}, K$^0_s$ \cite{k0}, $\phi$ \cite{phi}, 
$\Lambda$'s \cite{lambda}, $\Xi$'s \cite{xi} and $\Omega$'s \cite{omega}. These 
data samples are large enough to determine unambiguously the source parameters quoted 
in Table~1. 
On the other hand, for peripheral collisions, the full-phase-space multiplicities 
measured by NA49, i.e. $\pi$ \cite{per} and K \cite{per}, together with the number 
of wounded nucleons $N_W$ \cite{per}, are not sufficient to determine the source 
parameters. For an unambiguous fit, we need at least one ratio antibaryon/baryon, 
in order to determine the baryon-chemical potential. Once this is given, the ratio 
$N_W /\pi$ fixes the temperature and, in turn, the K/$\pi$ ratio fixes the strangeness 
suppression factor, $\gamma_S$. Antibaryon yields as a function of centrality have 
been measured by the NA57 Collaboration \cite{na57yields}. These yields cover only 
one unit of rapidity around midrapidity, however, so that some extrapolation to 
full-phase-space is necessary. To keep to a minimum the uncertainty related to the extrapolation,
we have used the ratio $\bar\Xi^+/\Xi^-$, because the rapidity distributions 
measured by NA49 in central collisions of both particle and antiparticle are gaussians 
to a very good approximation and an extrapolation is definitely
more reliable than for $\Lambda$'s. The procedure was as follows:
\begin{enumerate}
\item{} we have fitted the $\bar\Xi^+$ and $\Xi^-$ rapidity distributions
measured by NA49 in central collisions \cite{xi} to gaussians with normalization 
and width $\sigma$ left as free parameters and central value $y_0=0$ fixed;
\item{} assuming that the single gaussian shape and the ratio between their 
widths does not change as a function of centrality, we have estimated the yields in 
full phase space by rescaling the ratio $\bar\Xi^+/\Xi^-$ at midrapidity measured by 
NA57 by the ratio of the widths; in fact, for a single gaussian, the relation between 
midrapidity yield $dN/dy|_{y=0}$ and $N$ reads $ dN/dy|_{y=0} = N/\sqrt{2\pi} \sigma$.
\end{enumerate}
The quality of the gaussian fits is excellent and the widths of rapidity 
distributions turn out to be $0.860\pm 0.067$ for $\bar\Xi^+$ and $1.165 \pm 0.072$
for $\Xi^-$, implying a ratio $\sigma_{\bar\Xi^+}/\sigma_{\Xi^-} = 0.738 \pm 0.073$.
The obtained values of $\bar\Xi^+/\Xi^-$ in full-phase-space are quoted in Table~1 for the
different centrality bins. It should be remarked that the NA57 centrality bins 
corresponding to the $\Xi$'s yield values slightly different from those of NA49. 
However, this should imply a negligible error because the $\bar\Xi^+/\Xi^-$ is fairly 
constant over the whole centrality range.

The obtained source parameters for each centrality bin in Pb--Pb and for central 
C--C and Si--Si collisions are quoted in Table~1. The fit quality is good 
throughout. Temperature and baryon-chemical potential are essentially constant 
regardless of centrality and colliding system size. As expected, freeze-out temperatures are smaller than the temperature derived from $J/\psi$ opacity for $l<4$~fm, which refers to the initial stage of the fireball. 

Unlike $T$ and $\mu$, the strangeness
suppression parameter $\gamma_S$ varies considerably between the most central and 
the most peripheral bin in Pb--Pb. Also, $\gamma_S$ is significantly smaller in central 
C--C and Si--Si than in Pb--Pb collisions. These results are in agreement and 
confirm a previous analysis in Ref.~\cite{cley} where a slightly different data 
set was used. 

\begin{sidewaystable}
\begin{tabular}{|c|c|c|c|c|c|c|c|}
\hline 
                  & Pb--Pb          & Pb--Pb        & Pb--Pb        & Pb--Pb        & Pb--Pb          & C--C            & Si--Si \\ 
                  &(0-5)\%          &(5-12.5)\%     &(12.5-23.5)\%  &(23.5-33.5)\%  &(33.5-43.5)\%    & (0-15.3)\%      & (0.-12.2)\% \\	  
\hline  
 $N_P$            & 362$\pm$5.1     & 		    & 		    &		    &		      & 16.3$\pm$1.0	& 41.4$\pm$2.0	  \\
 $N_W$		  &		    &  281$\pm$10   &  204$\pm$10   &  134$\pm$10   &  88$\pm$10      &     	        &     		   \\
 $\pi^+$          & 619$\pm$35.4    &               &               &               &		      & 	        &	           \\		       
 $\pi^-$          & 639$\pm$35.4    &  519$\pm$25   &  394$\pm$19   &  290$\pm$14   &  191$\pm$9.1    & 22.4$\pm$1.6    & 56.6$\pm$4.1    \\		       
 K$^+$            & 103$\pm$7.1     &  80$\pm$9.1   &  54$\pm$5.5   &  35.5$\pm$3.7 &  20.5$\pm$2.3   & 22.2$\pm$1.6    & 57.6$\pm$4.1    \\ 		       
 K$^-$            & 51.9$\pm$3.6    &  45.1$\pm$4.5 &  29.7$\pm$3.0 &  20.2$\pm$2.0 &  11.8$\pm$1.3   & 2.54$\pm$0.25   & 7.44$\pm$0.74   \\ 		       
 K$^0_S$      	  & 81$\pm$4	    & 		    & 		    & 		    & 		      & 1.49$\pm$0.15   & 4.42$\pm$0.44   \\ 		       
 $\phi$           &  7.6$\pm$1.1    & 		    & 		    & 		    & 		      &       	        &		  \\ 		       
 $\Lambda$    	  &  44.9$\pm$6.4   & 		    & 		    & 		    & 		      & 0.178$\pm$0.024 & 0.66$\pm$0.085  \\		       
 $\bar\Lambda$	  &  3.74$\pm$0.47  & 		    & 		    & 		    & 		      & 1.37$\pm$0.29   & 3.88$\pm$0.58   \\		       
 $\Xi^-$          &  4.45$\pm$0.22  & 		    & 		    & 		    & 		      & 0.181$\pm$0.020 & 0.505$\pm$0.077 \\		       
 $\bar\Xi^+$      &  0.83$\pm$0.04  & 		    & 		    & 		    & 		      &  	        &  		  \\ 		       
 $\Omega$         &  0.59$\pm$0.13  & 		    & 		    & 		    & 		      &  	        &  		  \\ 		       
 $\bar\Omega$     &  0.26$\pm$0.067 & 		    & 		    & 		    & 		      &  	        &  		  \\ 		       
$\bar\Xi^+/\Xi^-$ &		    & 0.15$\pm$0.02 & 0.20$\pm$0.025& 0.20$\pm$0.025& 0.183$\pm$0.036 &   	        &   		  \\ 		       
\hline
 Parameter        &  & &  &  &  &    \\		
\hline   
  $T$ (MeV)       & 157.5$\pm$2.2   & 150.6$\pm$3.2  & 156.6$\pm$3.7  & 154.7$\pm$4.2  & 153.2$\pm$5.9  & 166.0$\pm$4.4  & 162.1$\pm$7.9   \\
  $\mu_B$ (MeV)   & 248.9$\pm$8.2   & 235.2$\pm$8.5  & 223.7$\pm$9.9  & 210$\pm$11     & 213$\pm$16     & 262$\pm$13	 & 260$\pm$18	   \\
  $\gamma_S$      & 0.842$\pm$0.038 & 0.709$\pm$0.071& 0.612$\pm$0.060& 0.542$\pm$0.054& 0.476$\pm$0.053& 0.547$\pm$0.041& 0.621$\pm$0.076 \\
  $\chi^2/$dof    & 22.5/9          & 0.29/1	     & 0.79/1	      & 0.86/1         & 0.59/1         & 4.1/4	   	 & 10.4/4    	   \\
\hline  
\end{tabular}
\caption{Particle abundances in heavy ion collisions and resulting values of thermodynamical parameters at chemical freeze-out . Top SPS energy, data from NA49 and NA57 (see text). Centrality bins are indicated for each collision.}
\end{sidewaystable}
An intriguing question about the strangeness production pattern in 
heavy ion collisions is whether a regular behaviour exists as a function of system 
size. Indeed, as can be seen in Fig.~3, $\gamma_S$ is larger in central 
light-ion collisions than in the most peripheral heavy-ion collisions, where it 
approaches the value found in pp collisions \cite{becaheinz,becman}. The search
of an appropriate scaling variable for $\gamma_S$ has been recently performed 
e.g. in Ref.~\cite{cley}, where a scaling of $\gamma_S$ as a function of 
the fraction of multiply struck participants in the Glauber model was pointed out. 
A different approach is to look for an underlying physical mechanism implying
the undersaturation of the strange phase space, with $\gamma_S$ thus becoming an 
effective parametrization of a more complicated physical picture: for instance, 
in Ref.~\cite{bkgms} a two-component model was proposed in which particle production 
stems from the superposition of a fully equilibrated hadron gas with $\gamma_S=1$ 
and single nucleon-nucleon collisions, where strangeness is strongly suppressed 
\cite{becaheinz}; H\"ohne {\it et al.} \cite{hoene} put forward a multi-cluster 
model where strangeness undersaturation is achieved by means of the so-called 
canonical suppression. The aim of explaining $\gamma_S$ in terms of a more 
fundamental physical scheme is certainly worth being pursued. However, it is a 
remarkable fact that, whatever its origin, strangeness undersaturation apparently 
scales with a pure geometrical variable such as $l$, as shown in Fig.~1.

\begin{figure}[ht]
\begin{center}
\epsfig{
height=7truecm, width=7.5truecm,
        figure=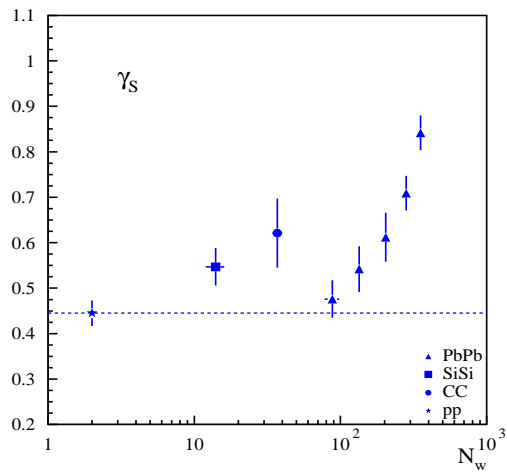}
\caption{\label{fig.3} \footnotesize The strangeness under-saturation factor, $\gamma_S$ as a function of the number of wounded nucleons in Pb-Pb (triangles), Si-Si (circle), C-C (box) and in p-p (star) collisions.}
\end{center}
\end{figure}

\section{Comparison of strangeness production and  anomalous $J/\psi$ suppression}
\label{}

In order to compare strangeness production to the $J/\psi$ anomalous suppression pattern  
one needs to find the $l$ values corresponding to the centrality bins defined 
by NA49. The subdivision is done on the basis of the number of projectile spectators 
measured in the forward calorimeter whose relevant 
fractions \cite{na49centr} are quoted in Table~1. The conversion to $l$ can be carried out by means of a 
Monte-Carlo simulation of the Glauber model.						    
Our Monte-Carlo simulation of the Glauber model follows the traditional scheme
\cite{misko}: at a given impact parameter, the number of collisions of each projectile
nucleon is randomly extracted from a Poisson distribution whose mean is the product
of the thickness function times the inelastic nucleon-nucleon cross-section. For
the target nucleons, the number of collisions undergone by each of them is randomly
extracted from a multinomial distribution constrained with the total number of 
collisions as determined in the previous step and whose weights are proportional
to the product of their relevant thickness function times the inelastic nucleon-nucleon 
cross-section. The thickness function has been calculated on the basis of a 
Woods-Saxon distribution:
\begin{equation}\label{ws}
  \frac{dN}{dr} = \frac{n_0}{1+{\rm e}^{(r-R)/d}} 
\end{equation}
with parameters quoted in Ref.~\cite{misko}:
$$
  n_0 = 0.17 \; {\rm fm}^{-3}; \;R= 1.12 A^{1/3}-0.86 A^{-1/3} \; {\rm fm};\;
d = 0.54  \; {\rm fm}.
$$
An issue in the calculation of $l$ is the definition of the radius $R$
if, as it was the case, a Woods-Saxon distribution is used. In fact, different 
definitions (e.g. the equivalent spherical radius, square mean radius etc.) result
in different values of $l$, though not dramatically. In the study of $J/\psi$ 
anomalous suppression, a value of 6.5 fm was adopted for the Pb radius, as mentioned 
above. To be consistent, we define the Carbon and Silicon radii as those encompassing 
the same fraction of the Woods-Saxon distribution as for the Pb nucleus with 6.5 
fm, namely 81\% with our parametrization (\ref{ws}). This leads to $R_{\rm C} = 
3.23$ fm and $R_{\rm Si} = 3.84$ fm. 
The resulting $l$ distributions for the NA49 different centrality bins in Pb--Pb 
collisions are shown in Fig.~4.

\begin{figure}[ht]
\begin{center}
\epsfig{
height=7truecm, width=7.5truecm,
         figure=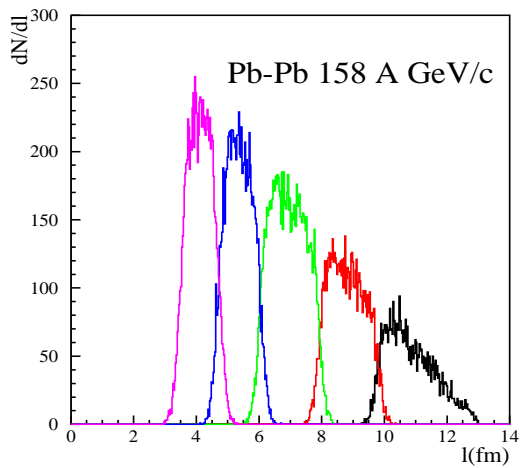}
\caption{\label{dndl} \footnotesize Probability distributions for $l$ in the different NA49 centrality bins in Pb--Pb 
collisions, as resulting from our Glauber Monte-Carlo calculation.}
\end{center}
\end{figure}

\section{Discussion and Conclusions}
\label{}

Fig.~1 gives convincing evidence that neither nuclear nor hadronic absorption 
can explain the drop in $J/\psi$ production at $l\simeq 4-5$~fm. In a Quark Gluon 
Plasma, the sudden drop indicated by Pb and In data would be attributed to the 
disappearance of $\chi_{c}$ and $\psi^{\prime}$\cite{Vogt&Satz}, which account 
for a sizable fraction of the observed $J/\psi$s in nucleon-nucleon collisions. 
To define precisely the expected signal in this case, however, one should have 
divided the experimental data by the $J/\psi$ absorption coefficient in the QGP 
medium. This may explain the residual slope displayed by the Pb points in Fig.~1, 
for $l>7$~fm. The limiting hadronic absorption estimated in Ref.~\cite{MPPR}, 
see Eq.~(\ref{abstot}), is  $\lambda^{-1}= \Sigma<\rho_i \sigma_i> \simeq 0.044$~fm$^{-1}$. 
To produce  a flat behavior  of the Pb points, one would need rather 
$\lambda^{-1}=0.13$. To our knowledge, there are no theoretical calculations 
available of the QGP absorption coefficient, which would be very useful for charmonium 
and bottomonium phenomenology at RHIC and LHC as well.

A second remark concerns $l$ dependence. One may wonder whether the energy density 
deposited in the fireball would be a more appropriate variable to use. Indeed, 
a plot of $J/\psi$ absorption vs. energy density estimated with the Bjorken formula
shows better scaling of Pb and In data. Conversely, the result for $\gamma_S$ looks like Fig.~3 and scaling is not achieved: the estimated
energy density in central Si--Si collisions is in fact significantly smaller than 
that in peripheral Pb--Pb collisions, whereas the converse is true for $\gamma_S$
values. Errors are still large, however, and more precise data as well as data 
from other A--A collisions are needed for better understanding.

In conclusion, we have pointed out that  $J/\psi$ suppression and strangeness enhancement are correlated to the transverse size of the interaction region in heavy ion collisions at SPS energies, $l$. We cannot offer an explanation in terms of fundamental physics and the present result is phenomenological. However it is interesting to note that $l$ determines the initial volume of the fireball and the energy density deposited by the collision. Collective phenomena like the ones we are addressing can take place most likely only when the hadron matter involved reaches a certain minimum volume and energy density and $l$ may thus be a good candidate for a critical lenght. 

The correlation we observe makes considerably stronger the case for the SPS being right at the onset of QGP formation. Further experimental investigations in this energy range are clearly called for, to elucidate the nature of the transition.

\section*{Acknowledgments}
	   								    
We are very grateful to M. Gazdzicki for illuminating discussions and invaluable help in handling NA49 data. We thank C. Lourenco, L. Ramello and E. Scomparin for discussing with us the NA50 and NA60 data and F. Antinori and H. Satz for interesting discussions.	   								    		   						

\end{document}